\documentclass[aps,prl,twocolumn,floatfix]{revtex4}
\usepackage{psfig}

\begin{document}

\title{
\parbox{30mm}{\fbox{\rule[1mm]{2mm}{-2mm}\Large\bf\sf PREPRINT}}
\hspace*{4mm}
\parbox{100mm}{\footnotesize\sf
Submitted for publication in {\it Phys. Rev. B} (2005) }\hfill
\\[5mm]
Mapping the optical properties of slab-type two-dimensional photonic
crystal waveguides}
\date{\today}

\author{Eric~Dulkeith}
\email[Corresponding author; e-mail: ]{fdulkei@us.ibm.com}
\affiliation{IBM T.~J.\ Watson Research Center, Yorktown Heights,
NY
  10598, USA}

\author{Sharee~J.\ McNab}
\affiliation{IBM T.~J.\ Watson Research Center, Yorktown Heights, NY
  10598, USA}

\author{Yurii~A.\ Vlasov}
\affiliation{IBM T.~J.\ Watson Research Center, Yorktown Heights, NY
  10598, USA}

\begin{abstract}

We report on systematic experimental mapping of the transmission
properties of two-dimensional silicon-on-insulator photonic
crystal waveguides for a broad range of hole radii, slab
thicknesses and waveguide lengths for both TE and TM
polarizations. Detailed analysis of numerous spectral features
allows a direct comparison of experimental data with 3D plane wave
and finite-difference time-domain calculations. We find,
counter-intuitively, that the bandwidth for low-loss propagation
completely vanishes for structural parameters where the photonic
band gap is maximized. Our results demonstrate that, in order to
maximize the bandwidth of low-loss waveguiding, the hole radius
must be significantly reduced. While the photonic band gap
considerably narrows, the bandwidth of low-loss propagation in PhC
waveguides is increased up to 125nm with losses as low as
8$\pm$2dB/cm.

\end{abstract}

\maketitle

\section{I. INTRODUCTION}

Two-dimensional photonic crystals (PhC) have gained considerable
interest and are regarded as a promising platform for dense
integration of planar photonic integrated circuits on a chip-scale
level~\cite{ref1,ref2,ref3,ref4,ref5,ref6,ref7,ref8,ref9,ref10,ref11,ref12,ref13,ref14,ref15,ref16,ref17,ref18,ref19}.
In particular, PhCs fabricated on silicon-on-insulator (SOI)
provide waveguide cross-sectional areas on the sub-micron scale
while maintaining single-mode operation due to high refractive
index
contrast~\cite{ref4,ref5,ref6,ref7,ref8,ref9,ref10,ref11,ref12,ref13,ref14}.
It has been shown that for triangular lattice membrane-type
PhCs~\cite{ref15,ref16,ref17}, consisting of a periodic array of
holes with a silicon slab thickness $h$ around 0.5$a$ and hole
radii of $\sim$ 0.35$a$ ($a$ is the lattice constant) the
bandwidth of the photonic band gap (PBG) can extend over most of
the 1.3 and 1.5$\mu$m telecommunications transmission bands.
Single-mode waveguiding can be realized by removing a single row
of holes in the periodic photonic lattice along the $\Gamma$-K
direction~\cite{ref16,ref17}. Mode confinement within the silicon
slab in the vertical direction is accomplished by index guiding,
while lateral confinement within the slab plane is defined by the
PBG~\cite{ref18}. This hybrid confinement limits the bandwidth for
intrinsically loss-less propagation to frequencies that are below
the light line cutoff. Above the light line cutoff the propagation
becomes lossy with mode-leakage out of the slab~\cite{ref6,ref10}.
Consequently, the bandwidth for intrinsically low-loss propagation
(LLP) of the waveguiding mode is restricted to only a fraction of
the PBG, e.g. in the order of only a few tens of
nanometers~\cite{ref7,ref8,ref9,ref10,ref13,ref14}. These insights
and many other theoretical investigations of light confinement in
planar PhCs and PhC waveguides have been examined in detail and
are well understood. In contrast, experimental studies of the
waveguiding properties of the slab-type PhC waveguides have been
hindered for a long time by difficulties in fabricating these
structures. One of the major challenges in the last decade was to
overcome extremely high transmission losses due to surface
roughness, structural irregularities and inefficient coupling,
which significantly complicated the interpretation of experimental
results. Recent achievements in optimizing the light coupling
efficiency into the PhC as well as fabrication improvements
reducing surface roughness have led to ultra low-loss PhC
waveguides thus enabling further spectroscopic
exploration~\cite{ref10,ref11,ref13,ref14}.

In this paper, we present detailed and comprehensive experimental
studies of the transmission properties of TE-like (even parity
with respect to the slab plane) and TM-like (odd parity with
respect to the slab plane) modes of two-dimensional membrane-type
PhC waveguides. The spectral properties are measured as a function
of various lattice parameters such as hole radius $r$, slab
thickness $h$, and the crystal length $l$. A survey of the applied
experimental techniques, fabrications methods and theoretical
calculations is given in Sec. II. In Sec. III we explore the
transmission properties of the PhC waveguides with structural
parameters optimized for a maximum bandgap. We show that for large
hole radii of $r/a$ $\sim$ 0.38 the bandwidth for LLP completely
disappears due to effective interaction with the band edge slab
modes of the PhC. The transmission properties of PhC waveguides
with small hole radii spanning from $r/a$=0.22-0.34 are
investigated in Sec. IV. We find, counter intuitively, that in
order to maximize the bandwidth of the waveguiding mode the hole
radius needs to be reduced to $r/a$ $\sim$ 0.2. Although this
optimization is accompanied by a significant decrease of the PBG,
the bandwidth for propagation is increased up to 125nm,
maintaining a reasonable trade-off between bandgap and propagation
bandwidth. The experimental findings are compared to 3D planewave
calculations using the MIT Photonic Band code~\cite{ref19} and
3D-Finite Difference Time Domain (FDTD)~\cite{ref20}. For the
propagation bandwidth-optimized parameters, we determine losses as
small as 8$\pm$2dB/cm, which is one of the lowest reported to
date. Sec. V. will deal with experimental results of transmission
spectra in TM polarization. After their comparison with photonic
band structures, in Sec. VI we will address the observed coupling
of even TE- with odd TE- and TM-like modes. Results are summarized
in Sec. VII.

\section{II. EXPERIMENTAL}

Devices were fabricated on lightly p-doped 200mm
silicon-on-insulator SOI wafers from SOITEC~\cite{ref24} with a Si
device layer thickness $h$ of 220nm on top of a 2$\mu$m buried
oxide layer (BOX). The processing is performed on a standard CMOS
fab line at the IBM Watson Research Center as described
elsewhere~\cite{ref10,ref11}. A 50nm thick oxide is deposited via
chemical vapor deposition to serve as a hard mask for etching. The
wafers are patterned by electron beam lithography (LEICA-VB6,
100keV, single writing field 400$\mu$m). The oxide hard mask is
opened and the silicon device layer dry is etched
$CF_{4}/CHF_{3}/Ar$ and $HBr$ chemistry, respectively. After
structure definition the BOX layer is selectively under-etched in
a buffered HF to form suspended membranes. PhC waveguides are
defined by omitting one row of holes from the PhC lattice in the
$\Gamma-K$ direction (W1-type waveguide). The light is coupled to
the PhC waveguide via a butt-coupled single-mode strip waveguide.
After definition of the PhC membrane the polymer inverted
spot-size converters are fabricated to achieve high coupling
efficiency as reported elsewhere~\cite{ref10,ref11}. Three
different wafers with slightly different slab thicknesses $h$ were
processed with PhC waveguides having nominally the same lattice
constant $a$=437.5nm, but different hole radii $r$=96, 140 and
165nm ($r/a$=0.22, 0.32, and 0.35, respectively). The PhC
waveguide length is varied in each writing field from 29$\mu$m to
2000$\mu$m for accurate loss determination. Four of such fields
grouped together constitute one experimental chip with the e-beam
exposure dose intentionally varied between each field. This allows
the phase space of explored lattice parameters to be further
broadened. In addition, proximity effects due to background
electron beam scattering induce further minor changes in the hole
diameter between the different PhC waveguides even within one
field. Such variation allows the exploration of a quasi-continuous
range of different lattice parameters, but at the same time
requires thorough statistical analysis of the particular PhC
waveguide on a given chip.

The lattice constant $a$, Si slab thickness $h$, and the hole
radii $r$ on a given sample are determined by statistical
evaluation of a number of SEM images obtained on a LEO SEM1560
model with identical imaging conditions (acceleration voltage,
focal distance, magnification, column and aperture alignment,
etc). Over 110 PhC waveguides were characterized in total. The
lattice constant was measured four times for each PhC waveguide
(total number of measurements $N$=440) yielding an overall value
of $a$=437.6nm with a standard deviation of only $\sigma$=0.93nm.

In order to check the variation of the hole radius along a single
PhC waveguide $N$=60 holes were measured at three different
locations along the longest 2000$\mu$m-long waveguide separated by
approximately 600$\mu$m. The results yield statistically identical
mean hole radii of 99.21nm at one end of the PhC waveguide,
99.18nm in the middle section, and 98.17nm at the other end with
standard deviations not exceeding 1.2nm. The overall mean hole
radius for the total of $N$=180 measurements was found to be
98.9nm with $\sigma$=1.12nm. For all other PhC waveguides the hole
radius was measured only six times in their middle section.
Although this small number imposes a less accurate estimation of
the mean value, however from the of $N$=180 measurements the
resulting standard error does not exceed 0.5$\%$ for $N$=6. The
slab thickness $h$ of the wafers was measured at nine different
positions along each sample. For the three processed SOI wafers
with devices having $r/a$=0.22, 0.32 and 0.35 the normalized
thickness of the Si slab was measured to be $h/a$=0.518, 0.507 and
0.546 respectively.

Measurements of transmission spectra of PhC waveguides are
performed using a broadband (1200-1700nm) LED source and an
optical spectrum analyzer (OSA) with 5nm spectral resolution. A
rejection ratio of over 30 dB between TE and TM polarizations is
achieved with the use of polarization maintaining, tapered and
lensed fibers and a polarization controller. The transmission
spectra of the PhC waveguides are normalized to the transmission
spectra of a corresponding single-mode reference strip waveguides
located in the same writing field near the PhC structures as
described elsewhere~\cite{ref10,ref11}.

The experimental results are compared with 3D plane wave
calculations performed using the MIT Photonic-Band
code~\cite{ref19}. The dielectric permittivity $\varepsilon_{Si}$
of the Si slab and $\varepsilon_{SiO2}$ of the oxide layer are
taken as 12.1 and 2.0, respectively. Values for the hole radii and
lattice constant are chosen as determined by SEM measurements. The
grid resolution (number of vectors in the unit cell of the PhC) of
the plane wave calculations is set to 16x16x16, which gives an
error in the eigenvalue convergence below $\sim$ 2$\%$ with
reasonable calculation time~\cite{ref19}. However, the limited
resolution imposes a drawback that very small changes in the slab
thickness $h$ are not well resolved especially for the TE-like
modes resulting in a step-like rather than a smooth dependence of
the calculated band structures on $h$. To overcome this issue, the
photonic crystal band structures for slab thicknesses with
$h/a$=0.4, 0.5, 0.6 and 0.7 are calculated and intermediate values
are interpolated. Interestingly, to achieve the best agreement
between experiment and theory all slab thicknesses needed to be
assumed 5-10$\%$ thinner than measured by SEM. This discrepancy is
consistent with a small, but noticeable thinning of the silicon
slab in the membrane region during the BOX etch.

All calculations of transmission spectra are performed with a 3D
finite difference time domain method (3D FDTD) using the Fullwave
software package~\cite{ref20}. Transmission through a 33 unit cell
long PhC W1 waveguide with 13 rows of holes on each side is
calculated with a 265x243x551 mesh (20nm grid size) for over 250000
time steps.

\begin{figure}[tb]
\begin{center}
\leavevmode \psfig{figure=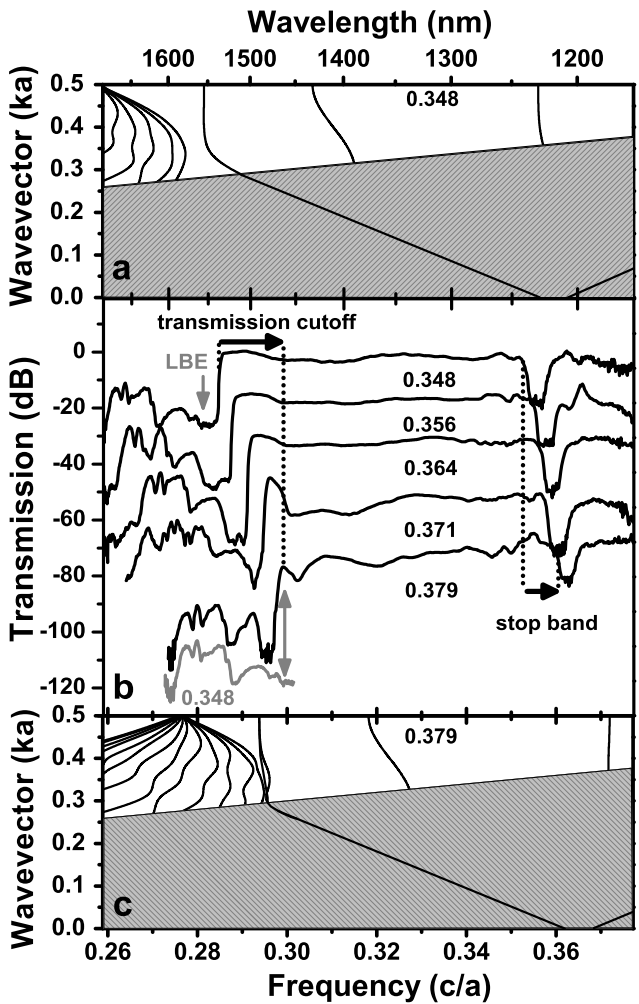,width=85mm}
\end{center}
\caption{Comparison of experimental TE transmission spectra of PhC
waveguides with large hole radii against 3D plane wave band
structure calculations. a) Band structure for a PhC waveguide with
a hole radius of $r/a$=0.348. b) Set of transmission spectra for
PhC waveguides, equal in length (l=29$\mu$m) but with increasing
hole radius $r/a$=0.348-0.379. The grey curve illustrates the
spectral merge of LBE and waveguiding onset for the PhC with
$r/a$=0.379 (slab modes for $r/a$=0.348 shifted with respect to
the normalized frequency for comparison with the modes for
$r/a$=0.379). c) Band structure for a PhC waveguide with a hole
radius of $r/a$=0.379.} \label{fig1}
\end{figure}

\section{III. PHC WAVEGUIDES WITH MAXIMIZED PHOTONIC BANDGAP (TE-LIKE
MODES)}

Historically most efforts on both three-dimensional and
two-dimensional PhCs, were focused on maximizing the bandwidth of
the PBG~\cite{ref15}. A large variety of different hole shapes and
crystal symmetries were explored in a planar 2D geometry. It was
shown~\cite{ref15} that for a 2D PhCs a triangular lattice of air
holes (hexagonal first Brillouin zone), produces the widest PBG
for a slab thickness of $\sim$ 0.5$a$ and hole radii of $\sim$
0.35$a$. Waveguiding can then be provided by omitting a row of
holes or increasing the hole radius in the channel. It is known
however, that the resulting mode suffers from light leakage out of
the plane for frequencies above the light line. Leakage losses
were measured~\cite{ref6,ref10} to be as high as 1500dB/cm, which
prohibits utilization of this wavelength range for useful
waveguiding. These inevitable diffraction losses limit the
bandwidth for intrinsic low-loss propagation (LLP) below the light
line to only a fraction of the
PBG~\cite{ref7,ref8,ref9,ref10,ref13,ref14}. Sec. III will first
report on the transmission properties of triangular 2D PhC
waveguides as a function of the hole radius that are chosen to
match the conditions for an optimized PBG ($r/a$=0.348-0.384).

Figure \ref{fig1} shows a series of transmission spectra of PhC
waveguides of identical length (29$\mu$m), but different hole
radii and compares them with photonic band structure calculations.
A number of spectral features of the TE-like modes can be directly
mapped to the band diagram. Figure 1a presents the photonic band
structure for the PhC waveguide with the smallest radius of
$r/a$=0.348. The onset of the waveguiding mode at wavelengths
around 1550nm, also known as the mode gap cutoff~\cite{ref7,ref8},
is clearly seen. In the corresponding measured spectrum, this
onset results in a sharp transmission cutoff around 1535nm
(Fig.1b, upper curve). The bandwidth for intrinsic LLP is strictly
bounded by this and the crossing of the waveguide mode with the
light line. Calculations show the light line cutoff is at $\sim$
1510nm (normalized frequency of 0.29$c/a$). Since the length of
the PhC waveguide is only 60 unit cells, the light line cutoff
manifests itself as only a weak kink at $\sim$ 1505nm in the
experimental spectra. The LLP bandwidth can be determined to be
$\sim$ 30nm. At longer wavelengths, small resonances are present
in the measured transmission spectrum. According to the photonic
band structure, these resonances are identified as a series of
slab modes below the lower photonic band edge (LBE). These modes
are delocalized over the whole slab with their fields mainly
concentrated in the dielectric material (dielectric
bands)~\cite{ref15,ref16}. At shorter wavelengths, around 1230nm,
the transmission reveals a distinct stop band. This stop band
opens due to the folding of the waveguiding mode at the edge of
the Brillouin zone at $k$=0. The upper band edge (UBE), which is
defined by the commencement of slab modes that are predominately
located in the holes (air bands)~\cite{ref15,ref16}, are not
observed in the current set of spectra since it lies at 1110nm
(0.40$c/a$), as determined by band structure calculations.

The series of transmission spectra in Fig.\ \ref{fig1}b shows that
as the hole radius increases, all characteristic spectral features
of the PhC waveguide undergo a shift to higher energies and is
confirmed by the photonic band structure in Fig.\ \ref{fig1}c
($r/a$=0.379). This fact can be easily explained by a decrease of
the effective refractive index of the slab. However, in addition,
the spectral shape of the transmission in the region of LLP
changes drastically. For the smallest hole radius the LLP
transmission region is characterized with a nearly flat top
profile. The propagation losses for waveguides with approximately
the same radius were measured before as small as 24$\pm$3dB/cm. In
contrast, for PhC waveguides with larger hole radii, first, the
flat top profile in the vicinity of the mode onset gradually
narrows and then completely disappears for $r/a$=0.379, second,
propagation losses in this region can be estimated to be as high
as 70dB/cm indicating the complete vanishing of the LLP for PhC
waveguides for large hole radii.

\begin{figure}[tb]
\begin{center}
\leavevmode \psfig{figure=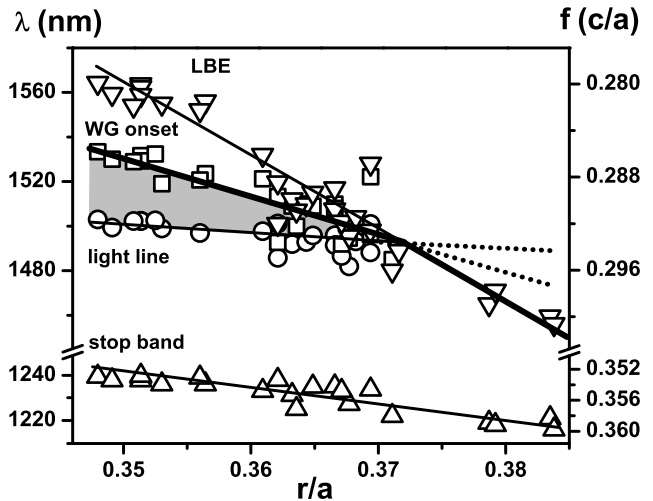,width=85mm}
\end{center}
\caption{Experimentally determined TE-map of spectral positions of
lower band edge (LBE), waveguiding onset, light line and stop band
as a function of the normalized hole radius for $r/a$=0.348-0.384.
The grey shaded area shows the frequency region for low loss
propagation (lines are guides for the eye).} \label{fig2}
\end{figure}

Interestingly, by analyzing Fig.\ \ref{fig1}, it can be seen that
with increasing hole radius the blueshift of the transmission
cutoff is far more pronounced than the corresponding blueshift of
the stop band. This is surprising since both stem from the same
waveguiding mode. To clarify this issue the experimental results
over a large number of PhC waveguides with different hole radii
are summarized in Fig.\ \ref{fig2}. The spectral positions of
waveguiding mode onset, stop band, light line cutoff and onset of
the LBE slab modes are plotted against normalized hole radii
(measured using a SEM on the very same waveguides). For this
particular range of hole radii the spectral position of the light
line is nearly constant. Since both, the mode onset and the stop
band are defined by the same PhC waveguiding mode, a change in
radius should lead to a parallel wavelength shift. The
experimental results for $r/a$$<$0.37 confirm this behavior. In
addition, the bandwidth for LLP continuously shrinks with
increasing the hole radius. On the other hand, the LBE exhibits a
much steeper spectral shift since the corresponding modes are
delocalized over the whole slab, thus being easily affected by
even small changes in the hole radius. The PhC waveguiding mode
instead is mainly localized in the center of the waveguide and is
less influenced by the surrounding holes. As a result, for hole
radii with $r/a$$>$0.37 the experimental transmission cutoff is no
longer defined by the waveguiding mode onset, but rather by the
onset of the slab modes at the LBE. This causes the LLP bandwidth
to completely vanish for large hole radii.

This experimental observation is confirmed by photonic band
structure calculations in Fig.\ \ref{fig1}a and Fig.\ \ref{fig1}c.
While for $r/a$=0.348 the waveguiding onset and the LBE are clearly
spectrally separated, for $r/a$=0.379 the LBE overlaps with the
waveguiding mode. Thus, light from the access strip waveguide can
couple not only effectively to the PhC waveguiding mode, but also to
the slab modes and hence leaks into the slab. This explains the
dramatic increase of the propagation losses. For PhC waveguides with
$r/a$$ > $0.37, it leads to a zero-bandwidth for low-loss
propagation. The results presented in this section demonstrate that
with structural parameters, chosen to maximize the PBG, the
bandwidth for LLP finally vanishes.

\section{IV. PHC WAVEGUIDES WITH MAXIMIZED LOW LOSS PROPAGATION BAND
(TE-LIKE MODES)}

Instead of optimizing the bandwidth of the PBG (a parameter rather
irrelevant for the implementation of PhC waveguides in integrated
optics), it is more meaningful to optimize the bandwidth of LLP.
From the analysis of the previous results in Sec. III we can
conclude that smaller hole radii are required for the optimization
of the LLP bandwidth. In Fig.\ \ref{fig2} we have seen that for a
decrease of the hole radius of only $\sim$ 5$\%$
($r/a$=0.365$\rightarrow$0.348) the LLP bandwidth significantly
broadens by more than a factor of two (13$\rightarrow$30nm). To
confirm this trend, this section will first deal with PhC
waveguides of only slightly smaller hole radii ($r/a$=0.32-0.34).
In the second part we will then proceed to investigate PhC
waveguides with very small holes ($r/a$=0.22-0.24).

\begin{figure}[tb]
\begin{center}
\leavevmode \psfig{figure=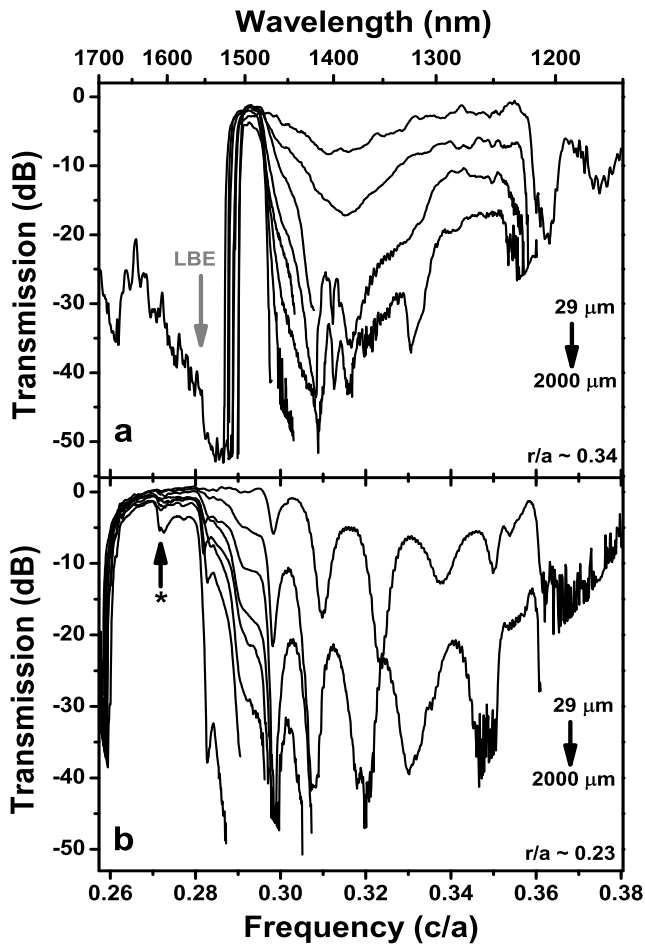,width=85mm}
\end{center}
\caption{Length series of TE transmission spectra of PhC
waveguides (29-2000$\mu$m) for two different hole radii: a) $r/a$
$\sim$ 0.34 and b) $r/a$ $\sim$ 0.23. The transmission dip marked
with an asterisk corresponds to the $x$-odd TE-like mode (see text
for details).}
\label{fig3}
\end{figure}

The experimental transmission spectra of PhC waveguides with
different lengths from 29 to 2000$\mu$m but approximately the same
hole radius ($r/a$$\sim$0.34) are shown in Fig.\ \ref{fig3}a. In the
same manner as done in Fig.\ \ref{fig1}, the comparison of
transmission spectra with photonic band structure calculations (not
shown) enables the identification of the spectral features of the
PhC waveguides. For the waveguide length of 29$\mu$m (upper curve),
the LBE is clearly visible in the long wavelength region at 1560nm.
The sharp transmission cut off, located at 1515nm corresponds to the
waveguiding onset and the light line cutoff is seen at 1480nm. At
shorter wavelengths, the transmission spectrum displays a stop band
at 1220nm. In fact, the spectra look very similar to the results of
Fig.\ \ref{fig1}b, however the PBG bandwidth is significantly
reduced by 25nm, while the bandwidth for LLP increases up to 35nm.
The results of Fig.\ \ref{fig3}a can be used to determine the
propagation losses in the LLP region. Losses at 1500nm are
calculated to be 20$\pm$3dB/cm; the method is described later in
this section (see also Ref.10).

Since the expansion of the bandwidth for LLP simultaneously takes
place with a shrinking of the PBG, one can conclude that there
must be a lower limit for the radius where the LLP bandwidth would
be restricted by the slab modes of the UBE. To find the upper
limit of the LLP bandwidth, PhC waveguides with significantly
reduced hole radii need to be investigated. Fig.\ \ref{fig3}b
shows the experimental results of a series of transmission spectra
of PhC waveguides with different lengths but again nearly the same
hole radii ($r/a$$\sim$0.23). Compared to Fig.\ \ref{fig3}a, the
size of the hole radius is reduced by over 30$\%$ and causes a
drastic change in the transmission spectra. Within the measured
spectral range, no LBE is detected, since it has redshifted to
1730nm. Furthermore, the stop band has also disappeared from the
spectra. On the very edge of the transmission spectra, around
1695nm, a sharp transmission cut off is present, and for shorter
wavelengths is followed by a region of flat transmission and
relatively low losses. At wavelengths shorter than 1550nm, the
transmission spectra are now dominated by very strong resonances
covering a broad bandwidth.

\begin{figure}[tb]
\begin{center}
\leavevmode \psfig{figure=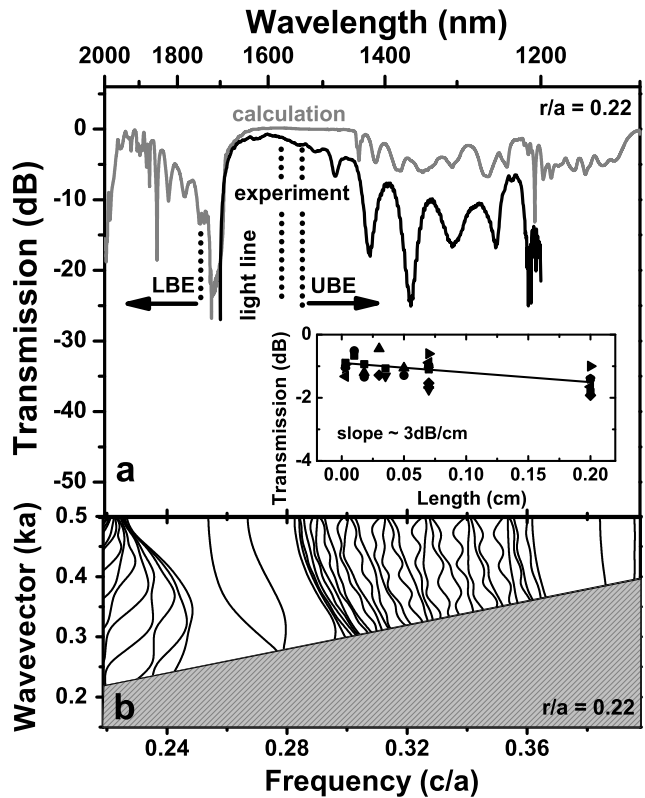,width=85mm}
\end{center}
\caption{a) Experimental TE transmission spectra (black line) of a
29$\mu$m PhC waveguide with $r/a$=0.22 compared with 3D FDTD
calculations (grey line). Inset: Absolute transmission at 1650nm
of PhC waveguides as a function of the length. From the slope, the
losses of the whole photonic circuit (PhC and strip waveguides)
are determined as 3$\pm$2dB/cm. This corresponds to 8$\pm$2dB/cm
loss for the PhC waveguide (see text for details). b) 3D plane
wave band structure calculations with the same structural input
parameters as for the FDTD calculations in a).} \label{fig4}
\end{figure}

To understand the origin of these resonances and to interpret the
experimental transmission spectra, Fig.\ \ref{fig4}a compares the
transmission spectrum of a 29$\mu$m PhC waveguide with $r/a$=0.22
against 3D FDTD calculations. For calculations the thickness of
the slab was defined as $h/a$=0.52. The FDTD calculated spectrum
is in good agreement with the experimental transmission spectrum.
The difference in amplitude is due to the shorter waveguide length
used in the calculations (33 unit cells or 14.4$\mu$m). The
comparison of calculated and experimental transmission spectra
allows identification of the measured transmission cutoff around
1700nm as the onset of the waveguiding mode, The onset is followed
by numerous resonances starting at 1725nm. In direct analogy with
the interpretation of the experimental results shown in Fig.\
\ref{fig1}b and Fig.\ \ref{fig3}a, these resonances can be
attributed to modes below the LBE.

The calculated transmission also confirms the presence of
additional resonances between 1200 and 1500nm. However, from the
FDTD spectrum alone it is not possible to assign their origin. To
address this issue Fig.\ \ref{fig4}b presents photonic band
structure calculation consistent with the structural input
parameters in Fig.\ \ref{fig4}a. The strong resonances clearly
match with slab modes whose spectral onset defines the UBE. The
UBE, not visible in the previous spectra of Fig1b and Fig3a for
large hole radii, has now redshifted over 400nm, almost spectrally
matching the position of the light line cutoff. This huge spectral
shift of the UBE is far more pronounced than the shift of the LBE.
As mentioned before, UBE (air)-modes are mainly located in the
holes of the PhC and thus exhibit a strong dependence on hole
radius. In the experimental transmission spectrum of Fig.\
\ref{fig4}a, a kink around 1570nm indicates where the waveguiding
mode crosses the light line. This becomes more visible for longer
waveguide lengths as shown in Fig.\ \ref{fig3}b (note this is for
$r/a$$\sim$0.23). The bandwidth for LLP in Fig.\ \ref{fig4}a. is
now close to 130nm, an increase of nearly 300$\%$. This very broad
bandwidth is confirmed by comparison with the corresponding
photonic band structure in Fig.\ \ref{fig4}b.

Losses within the LLP band are determined from transmission
spectra of PhC waveguides with different length. The inset in
Fig.\ \ref{fig4}a presents the absolute transmission values
measured at 1650nm. The slope of the transmission attenuation
gives 3$\pm$2dB/cm. Note that this number corresponds to the total
transmission losses through the PhC waveguide and access strip
waveguides, rather than characterizing the losses in the PhC
waveguide alone. While the length of the PhC waveguides and their
corresponding losses is increasing, at the same time the length of
the strip waveguides is shortened to keep the total length of the
photonic circuit on a chip constant at 4.6mm. Correspondingly, as
it was pointed out in Ref.10, for a correct determination of
propagation losses in the PhC waveguide it is necessary to account
for losses in the bare strip waveguide. Losses from TE-like modes
in analogous strip waveguides with a 460x220nm cross-section have
been determined recently as 3.6dB/cm and 5dB/cm at 1500nm and
1650nm respectively, as reported elsewhere~\cite{ref21}.
Correspondingly, the effective loss of the PhC waveguide alone at
1650nm can be estimated as 8$\pm$2dB/cm. To our knowledge, this is
one of the lowest loss numbers reported for SOI PhC waveguides.
There is a remarkable dependence on the hole radius. Losses
decrease from 24$\pm$3dB/cm for $r/a$=0.365 (Ref.10) to
20$\pm$3dB/cm for $r/a$=0.34 and to 8$\pm$2dB/cm for $r/a$=0.23.
It can be argued that the surface area of the holes is
significantly decreased in the waveguides with smaller radii, thus
scattering losses due to sidewall roughness are becoming less
pronounced. In addition, the effective waveguide width increases
with decreasing hole radius which can further reduce losses.

\begin{figure}[tb]
\begin{center}
\leavevmode \psfig{figure=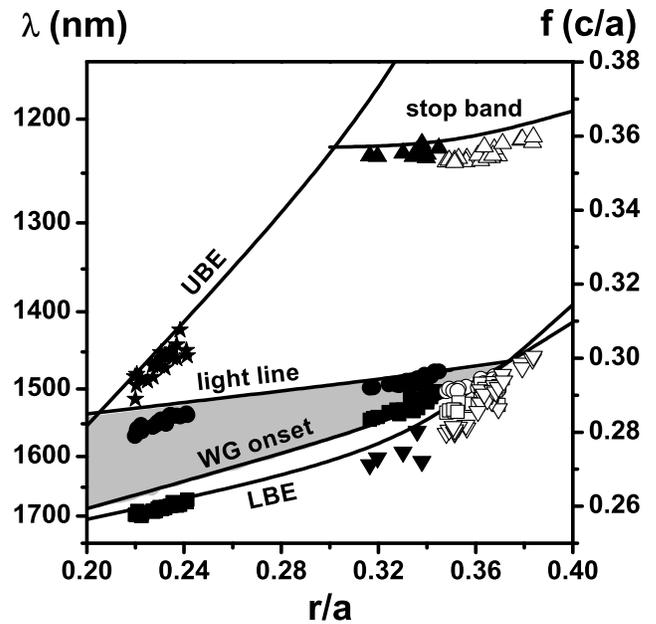,width=85mm}
\end{center}
\caption{Experimentally derived TE-map depicting spectral positions
of stop band (triangles up), UBE (asterisks), light line (circles),
WG onset (squares) and LBE (triangles down) as a function of the
hole radius $r/a$=0.22-0.38. Solid lines represent 3D plane wave
band structure calculations with input parameters determined from
SEM measurements for hole radii with $r/a$=0.32-0.34.}
\label{fig5}
\end{figure}

Figure \ \ref{fig5} summarizes the experimental results of the PhC
waveguides for all measured hole radii. The spectral positions of
the TE spectra such as stop band, UBE, LBE, light line and
waveguiding onset are shown as a function of the normalized hole
radius (open shapes correspond to data from Fig.\ \ref{fig2}). The
experimental results are compared to 3D planewave calculations
(solid line) for the slab thickness of $h$ taken as 0.473$a$. This
corresponds to the hole size series ranging from $r/a$=0.32-0.34.
Good agreement is found overall between experiment and
calculation. The results in Fig.\ \ref{fig5} clearly demonstrate
the broadening of the bandwidth for LLP, which is accompanied by a
simultaneous shrinkage of the PBG with decreasing hole radii. A
closer look shows that the set of data points for the three
different size series have a minor spectral displacement with
respect to each other. This step-like behavior can be explained by
the different slab thicknesses $h$ of the SOI wafers as measured
with a SEM (see Sec. II).

As can be seen from the planewave calculations in Fig.\
\ref{fig5}, the maximum LLP with over 150nm bandwidth is achieved
for hole radii $r/a$=0.207. Here the PBG bandwidth is reduced to
160nm, however the LLP region occupies nearly 94$\%$ of the
bandwidth. Apparently, further decrease of the radii would shrink
the LLP bandwidth again, as the slab modes of the UBE would merge
into the LLP.

\section{V. TRANSMISSION OF THE TM-LIKE MODES IN PHC WAVEGUIDES}

It is well known that a PBG for the TM-like modes does not exist for
SOI PhC slabs~\cite{ref15}. However, it is important to characterize
the transmission characteristics of PhC waveguides for TM-mode as
well. Firstly, measurements of TE- and TM-like modes provide good
verification and validation of theoretical fitting. With a set of
only three fitting parameters (hole radius, refractive index and
slab thickness) all spectral features found in both, TE- and
TM-spectra, have to simultaneously match the band structure
calculations. Secondly, and more importantly, due to possible
imperfections and errors in the experimental structures, TM-like
modes in principle may interact with the TE-like modes of interest
and cause additional propagation losses~\cite{ref11}.

\begin{figure}[tb]
\begin{center}
\leavevmode \psfig{figure=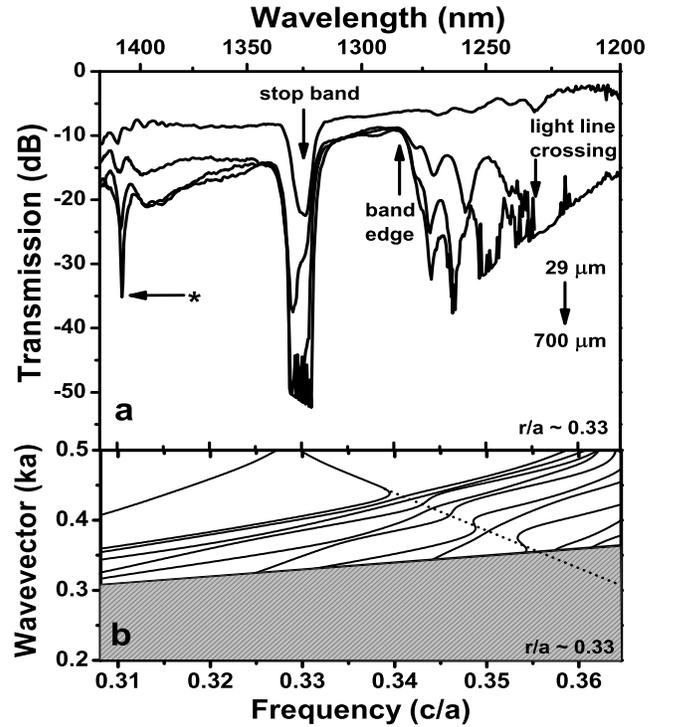,width=85mm}
\end{center}
\caption{a) Length series of TM transmission spectra of PhC
waveguides (29-700$\mu$m) with hole radii $r/a$$\sim$0.33. The
transmission dip marked with an asterisk corresponds to the $x$-odd
TE-like mode (see text for details). b) 3D plane wave band structure
calculations. The dotted line indicates the fundamental TM-like
waveguiding mode.}
\label{fig6}
\end{figure}

Measured transmission spectra of TM-like modes are shown in Fig.\
\ref{fig6}a. The hole radius of the PhC waveguides is
$r/a$$\sim$0.33. The length of the different photonic crystals range
from 29 to 700$\mu$m. All transmission spectra exhibit a strong dip
at around 1325nm, as well as multiple strong resonances starting at
1280nm and extending up to the transmission cutoff at 1235nm. As
before, 3D plane wave calculations of the photonic band structure
are helpful in interpretation of the spectral features (Fig.\
\ref{fig6}b). A dotted line indicates the fundamental TM-like
waveguiding mode. Both in- and out-of-plane confinement of the mode
is purely index guided, hence showing a nearly linear dispersion.
However, the mode is still strongly confined in the PhC waveguide
and provides most of the transmission. The dip at 1325nm can be
identified as a stop band that arises from the folding of the
fundamental mode at the edge of the Brillouin zone. This fundamental
mode is crossing the light line at around 0.355$c/a$, which explains
the appearance of the cutoff at 1235nm. Multiple resonances between
1280nm and 1235nm are arising from the coupling of the fundamental
mode with numerous slab modes. Such behavior is very analogous to
the mode mixing mechanism observed recently in SOI double-trench
waveguides~\cite{ref11}. The spectral region between the stop band
at 1325nm and the onset of the resonances at 1280nm is characterized
by relatively low-loss propagation. Measurements at 1300nm indicate
losses are as small as 16$\pm$3dB/cm.

\begin{figure}[tb]
\begin{center}
\leavevmode \psfig{figure=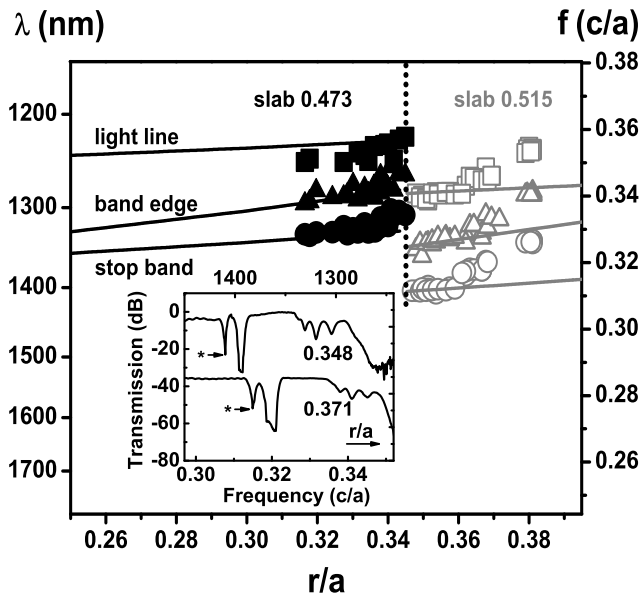,width=85mm}
\end{center}
\caption{Experimental TM-map of the spectral positions of light line
(squares), band edge (triangles) and stop band (circles) as a
function of the hole radius $r/a$=0.32-0.38. Solid lines represent
3D plane wave band structure calculations (black and grey) with
input parameters according to the experimental SEM results for hole
radii of $r/a$=0.32-0.34 or $r/a$=0.35-0.38 respectively. Inset: TM
spectra of two PhC waveguides, identical in length (29$/mu$m), but
with different hole radii (0.348$a$ and 0.371$a$). The transmission
dips marked with an asterisk correspond to the $x$-odd TE-like mode
(see text for details).} \label{fig7}
\end{figure}

The results of the analyzed TM transmission spectra for hole radii
from $r/a$=0.32-0.38 are summarized in Fig.\ \ref{fig7}. The solid
lines are the result of photonic band structure calculations. For
the two size series, a slab thickness of $h/a$=0.473 or 0.515 was
chosen as an input parameter. These are the same values as for the
corresponding TE-like photonic band structures in Fig.\ \ref{fig6}
and Fig.\ \ref{fig1}, respectively. The calculations are in good
agreement with the experimental results. The inset of Fig.\
\ref{fig7} shows the TM spectra of two waveguides, identical in
length (29$\mu$m), but different hole radii (0.348$a$ and
0.371$a$). The shift of the transmission spectra for differently
chosen hole radii is clearly visible. However, comparison with the
corresponding TE spectra of the identical waveguides (see Fig.\
\ref{fig1}b) shows that the TM modes exhibit a less pronounced
dependence on the hole radius. In contrast to this, comparing the
map of TM-like modes in Fig.\ \ref{fig7} to the map of TE-like
modes in Fig.\ \ref{fig5}, the spectral shift between the series
of different SOI wafers is far more pronounced. Such behavior is
not unexpected since TM-like modes are polarized perpendicular to
the slab plane and are therefore strongly influenced by a change
in the slab thickness and less by the hole radius.

\section{VI. COUPLING BETWEEN EVEN AND ODD MODES}

In principle, a W1 SOI waveguide is not single-moded over the
entire bandwidth of the PBG. It is known that it can support two
TE-like modes (both even with respect to the $xy$-plane in the
slab) possessing different symmetries with respect to the
$zx$-plane vertically bisecting the slab along the waveguide
direction ($x$-axis). All results presented so far correspond to
the in-plane $x$-even TE-like mode. This mode is of most interest
for integrated
photonics~\cite{ref4,ref5,ref6,ref7,ref8,ref9,ref10,ref11,ref12,ref13,ref14}.
The TE-like mode with odd $zx$-symmetry is visible in all band
structure calculations (e.g. Fig.\ \ref{fig1}a at $c/a$=0.31).
However, the excitation of this $x$-odd TE-like mode with the
$x$-even TE mode in the access strip waveguide is prohibited due
to symmetry restrictions. The orthogonality of the $x$-even and
$x$-odd TE-like modes in the PhC waveguides also prohibits any
interaction. Correspondingly, the odd mode is usually not observed
in transmission spectra and its properties are regarded as
irrelevant to waveguiding in PhC waveguides.

Indeed, in the experimental TE spectra of PhC waveguides with large
hole radii (Fig.\ \ref{fig1}b) no evidence of this mode at 1430nm
($r/a$=0.348) or 1390nm ($r/a$=0.379) is observed. TE transmission
spectra of PhC waveguides with small hole radii (Fig.\ \ref{fig3}b)
are also well explained by the photonic band structure and FDTD
calculations, as discussed above. However, there is another spectral
feature present in these spectra, which does not fit into this
picture of forbidden mode coupling. At wavelengths around 1600nm, a
distinct break down within the LLP region is observed (marked with
an asterisk in Fig.\ \ref{fig3}b). Its attenuation gradually
increases as a function of the PhC waveguide length. As seen from
the band structure calculations in Fig.\ \ref{fig4}b, in this
spectral range no other modes except the in-plane $x$-odd TE-like
mode are present. Since no indication of the $x$-odd TE-like mode is
visible in the spectra of short waveguides (Fig.\ \ref{fig1}b, Fig.\
\ref{fig3}b), we can conclude that this is not an interface
phenomenon but rather the $x$-even TE-like mode, being excited from
the strip waveguide and gradually transferring energy to the $x$-odd
mode. The field energy, once transferred, cannot couple efficiently
into the output strip waveguide due to symmetry restrictions and
therefore dissipates in the slab. Correspondingly, the $x$-even
TE-like mode exhibits additional propagation losses, which can be
estimated from Fig.\ \ref{fig3}b as 20dB/cm at the center of the dip
at around 1600nm. This weak efficiency of mode coupling explains the
absence of the $x$-odd mode in Fig.1b since the presented PhC
waveguides are only 29$\mu$m in length and the resulting losses
would only be $\sim$ 0.5dB. Furthermore, the frequency range where
both modes coexist is located far above the light line
($c/a$$\sim$0.32 in Fig.\ \ref{fig1}c) where diffraction losses due
to out-of-plane leakage dominate.

In general, the presence of both modes in the experimental spectra
indicates that the perfect symmetry of the PhC lattice with
respect to $zx$-plane is somehow relaxed and the interaction
between the modes can become possible. Two different mechanisms
can in principle be considered that might cause the violation of
symmetry in the otherwise highly symmetric (001) plane of the SOI
wafer: structural irregularities in the PhC lattice or in-plane
optical anisotropy.

Asymmetry in the waveguide can in principle be introduced by
structural imperfections of the PhC lattice. Sidewall surface
roughness was measured to be only a few nm~\cite{ref21} and
therefore is believed to be an unlikely candidate. Small lateral
offsets of the crystal can occur due to imperfect field and
sub-field stitching when devices are defined with electron-beam
lithography. Offsets in the $y$-axis would result in an
$xz$-asymmetry of the PhC waveguide and could be a possible cause
of mode-mixing. A field (with dimensions of 400$\mu$m) usually has
a stitch error $<$10-20nm while sub-fields ($\sim$6$\mu$m) are
typically even smaller, however the impact of stitching error
requires further systematic investigation.  A further possible
source of asymmetry is stress in the Si film. While Si is a cubic
crystal, and should therefore be optically isotropic, small
birefringence ($\triangle$$n$$\sim$$10^{-6}$) has been
reported~\cite{ref22}. Significant birefringence can be induced by
stress, and is a known problem for SOI rib
waveguides~\cite{ref23}. However, for released Si membranes
fabricated on Unibond SOI wafers~\cite{ref24} it is difficult to
expect strain higher than a few tens of MPa. Further investigation
is necessary to confirm the origin of the observed mode mixing.

Interestingly, the spectral signature of the $x$-odd TE-like mode
can also be identified in TM transmission spectra. For example, the
data in Fig.\ \ref{fig6}a, which shows TM transmission spectra of
PhC waveguides with $r/a$=0.33, exhibit a pronounced dip around
1410nm (again marked with an asterisk). Similar dips are also
observed in the TM transmission spectra of PhC waveguides of larger
hole radii (Inset of Fig.\ \ref{fig7}). The corresponding photonic
band structure of the TM modes in Fig.\ \ref{fig6}b cannot explain
this spectral feature. Instead, its spectral position exactly
coincides with the $x$-odd TE-like mode at the edge of the Brillouin
zone at $k$=0.5 (Fig.\ \ref{fig1}a,c and Fig.\ \ref{fig4}b).

\begin{figure}[tb]
\begin{center}
\leavevmode \psfig{figure=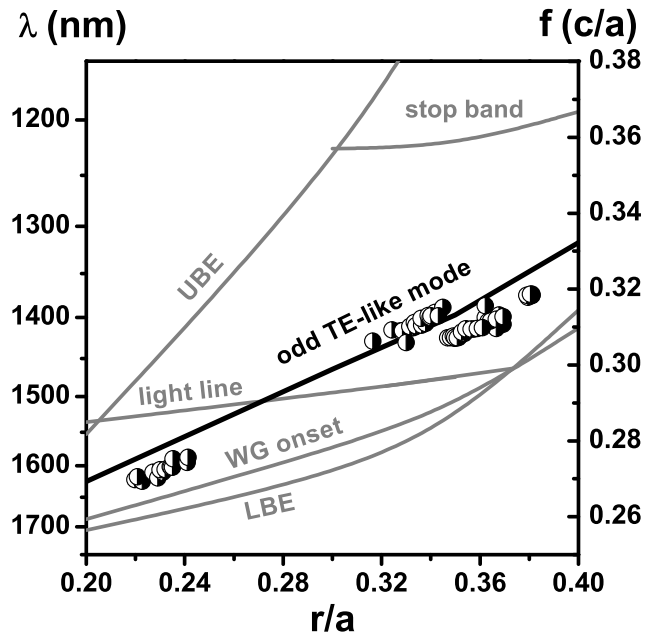,width=85mm}
\end{center}
\caption{Experimental map of the $x$-odd TE-like waveguiding mode as
a function of the hole radius. Solid lines represent 3D plane wave
band structure calculations with input parameters according to SEM
measurements for hole radii of $r/a$=0.32-0.34.  The black line
plots the $x$-odd TE mode while the grey lines represent the UBE,
light line, WG onset (even TE mode), stop band and LBE.}
\label{fig8}
\end{figure}

The spectral positions of these dips in both, TE and TM spectra
are plotted in Fig.\ \ref{fig8}. Data points for $r/a$=0.22-0.24
are taken from TE spectra and for $r/a$=0.32-0.38 from TM spectra.
The solid line represents the calculated dependency of the $x$-odd
TE-like waveguiding mode on the hole radius for a slab thickness
of $h/a$=0.473. In analogy to Fig.\ \ref{fig5}, the spectral
step-like shift between the different size series again follows
the change in the slab thickness. Remarkably all data points for
both, TE and TM spectra are in good agreement with plane wave
calculations for the $x$-odd TE-like mode over the entire range of
hole radii.

Just as for the TE-like transmission spectra, the attenuation at
the center of the dip in the TM spectra gradually increases with
waveguide length. We can therefore again exclude any significant
contribution of energy transfer between the modes already at the
interface between the strip waveguide and the PhC. The energy
transfer between the TM-like and TE-like $x$-odd mode is taking
place gradually while the light is propagating along the
waveguide.

\section{VII. SUMMARY}

In this paper we have performed an extensive investigation of the
transmission properties of SOI-type PhC W1-waveguides as a
function of the hole radius, slab thickness and the crystal length
for both TE and TM polarizations. Our experimental results reveal
that for structural parameters optimized for maximization of the
PBG, the bandwidth for low-loss propagation below the light line
cutoff completely vanishes for hole radii $r/a$$>$0.37. This is
mainly caused by a tremendous spectral shift of the lower band
edge slab modes, which are mostly localized in the slab
surrounding the holes and are strongly influenced by a decrease of
the average refractive index of the slab. PhC W1-waveguides with a
maximized PBG are therefore unsuitable for integrated photonics
applications. The experimental findings for PhC waveguides
possessing small holes with only $r/a$$\sim$0.2 demonstrate that
the bandwidth for transmission can be maximized to almost 130nm
while still maintaining a sufficient PBG. Losses for these devices
are reduced to only 8dB/cm presumably due to a significantly lower
surface area of the holes sidewalls. The experimental transmission
spectra for both TE and TM polarizations show very good agreement
with calculated photonic band structures and 3D-FDTD transmission
spectra.

\section{ACKNOWLEDGMENTS}

The authors gratefully acknowledge useful discussions with N. Moll
(IBM Zurich Research Lab) and G.L.Bona (IBM Almaden Research lab)
and the contributions of the Microelectronics Research Laboratory
staff at the IBM T. J. Watson Research Center.

\end{document}